\documentclass[prl,twocolumn,floatfix,superscriptaddress]{revtex4-1}
\usepackage{dcolumn,amsmath}
\usepackage[dvips]{graphicx}
\usepackage{bm}
\usepackage{hyperref}

\begin{document}
\title{Centrifugal correction to hyperfine structure constants in the ground state of lead monofluoride, PbF}
\author{A.N.\ Petrov}\email{alexsandernp@gmail.com}
\author{L.V.\ Skripnikov}
\author{A.V.\ Titov}
\affiliation
{Petersburg Nuclear Physics Institute, Gatchina,
             Leningrad district 188300, Russia}
\affiliation{Division of
Quantum Mechanics, St.Petersburg State University, 198904, Russia}
\author{R. J.\ Mawhorter}
\affiliation
{Department of Physics and Astronomy, Pomona College, Claremont, California 91711, USA}
\begin{abstract}
The sensitivity of the PbF molecule to the electron electric dipole moment has motivated detailed microwave spectroscopy. 
Previous theoretical approaches cannot fully explain the spectra. In turn, the explanation from ``first principles'' is very important both for molecular theory and for confirmation of
the correctness of the interpretation of experimental data obtained with high precision.
All of these issues are decisively addressed here.
We have determined centrifugal correction parameters for hyperfine structure constants, both on lead and fluorine nuclei, of the $X^2\Pi_{1/2}$ state of lead monofluoride. These parameters were obtained by fitting experimentally observed pure rotational transitions and from  {\it ab~initio} calculations. 
We show that taking this correction into account is required to reproduce the experimental transition energies obtained in [Phys.\ Rev.\ A {\bf 84}, 022508 (2011)].

\end{abstract}

\maketitle



Interest in the theoretical study of PbF is  motivated primarily by the proposed experiments to search for the simultaneous violation of time-reversal invariance (T) and space parity (P) \cite{Kozlov:87, Dmitriev:92, Shafer-Ray:06, Shafer-Ray:08E, Baklanov:10}. An experiment to search for the P,T-odd property of the electron, i.e.\ its electric dipole moment or eEDM, using trapped PbF molecules is currently under consideration \cite{Shafer-Ray:06}. Another experiment using PbF to measure the P-odd anapole moment has also been recently suggested \cite{Alphei:11}. The hyperfine structure (HFS) constants for the ground state of PbF are known with high accuracy \cite{Mawhorter:11}. The hyperfine structure of the rotational levels of such a molecule is usually described by an effective spin-rotational Hamiltonian. Parameters of this Hamiltonian can be obtained by fitting experimentally observed transitions. The accuracy of the measured transitions in \cite{Mawhorter:11} was so high that some terms new to the traditionally used spin-rotational Hamiltonian were introduced (see eq.~(5) in \cite{Mawhorter:11}) to describe them. The nature of these parameters is currently unknown.
It is unclear how they can be extracted from electronic structure calculations. The goal of the present paper is to reproduce the rotational spectra of the $X^2\Pi_{\pm1/2}$ state obtained by Mawhorter {\it et al.}~\cite{Mawhorter:11} from ``first principles'', i.e.\ without any artificial parameters. 
This is important both for molecular theory and for confirmation of
the correctness of interpretation of experimental data obtained with high precision. 
To verify the nature of the new terms introduced in the present paper, the parameters of the spin-rotational Hamiltonian are obtained in two different ways, from fitting the experimentally observed transitions and from {\it ab~initio} calculations.

In the present paper the hyperfine structure of rotational levels of the $^2\Pi_{\pm1/2} $ electronic state is obtained by numerical diagonalization of the molecular Hamiltonian (${\rm \bf H}_{\rm mol}$) on the basis set of the electronic-vibrational-rotational wavefunctions

\begin{eqnarray} 
\chi(R)_J\Psi_{^2\Pi_{\pm1/2}}\theta^{J}_{M,\pm 1/2}(\alpha,\beta)U^{\rm F}_{M_I}U^{\rm Pb}_{M_I}\ , \\
\chi(R)_J\Psi_{^2\Pi_{\pm3/2}}\theta^{J}_{M,\pm 3/2}(\alpha,\beta)U^{\rm F}_{M_I}U^{\rm Pb}_{M_I}\ .
\label{basis}
\end{eqnarray}
The vibrational wavefunction is $\chi(R)_J$ and it is determined by the equation

\begin{eqnarray}
\left(-\frac{\hbar^2d^2}{2M_rdR^2} + \frac{\hbar^2}{2M_r}\frac{J(J+1)}{R^2} + V(R)\right )\chi_J=
E_J\chi_J\ .\qquad
\label{Shr}
\end{eqnarray}
%
Here $\theta^{J}_{M,\Omega}(\alpha,\beta)=\sqrt{(2J+1)/{4\pi}}D^{J}_{M,\Omega}(\alpha,\beta,\gamma{=}0)$ is the rotational wavefunction and $U^{F,Pb}_{M_I}$ are the nuclear spin wavefunctions on F and Pb for $^{207}$Pb$^{19}$F. For $^{\rm even}$Pb$^{19}$F, the nuclear spin wavefunction on Pb $U^{\rm Pb}_{M_I}=1$. The scheme of the calculation is similar to that in \cite{Petrov:11}, although here we additionally introduce J-dependence for the vibrational wave functions. As one can see below, this is required to reproduce the experimental transition energies of the $^2\Pi_{\pm1/2} $ state.

\section{Electronic structure calculations}

A series of {\it ab~initio} calculations was performed in the present paper. To compute the hyperfine structure constants of the PbF molecule a scheme which combines the generalized relativistic effective core potential (GRECP) approach \cite{Mosyagin:10a,Titov:99} with the non-variational restoration procedure \cite{Titov:06amin} was used. The ``two-step'' approach has recently been used in \cite{Kudashov:13, Baklanov:10, Petrov:07a, Skripnikov:09, Skripnikov:11a, Petrov:11} for the calculation of other ``core properties'' (such as effective electric field on electron, Schiff moment enhancement factor etc.) in molecules and atoms.

To compute the hyperfine structure constants for the ground and excites states of PbF, fully-relativistic Fock-space coupled cluster code with single and double cluster amplitudes (FS-RCCSD) \cite{MolRCCSD, Kaldor:04ba} was applied. 
In this calculation both the valence and outer-core ($5s^25p^65d^{10}6s^26p^2$) electrons of Pb and all the electrons of F (i.e.\ 31-electrons in total) were treated explicitly while the inner-core electrons of Pb were excluded using the GRECP method.

To calculate the Pb hyperfine parameters, a basis set consisting of $8s$, $7p$, $4d$, $1f$ and $1g$ ([8,7,4,1,1])  contracted Gaussian basis functions on Pb and $5s$, $4p$, $2d$, $1f$ ([5,4,2,1]) functions on F was used. To calculate the F hyperfine parameters the basis set on Pb was reduced to [8,6,3,1], while the basis set on F was increased to [7,6,3,1]. The original set of basis functions on Pb was taken from a previous paper \cite{Isaev:00}, while for F the ANO-L atomic basis function from the MOLCAS 4.1 library \cite{MOLCAS} was used.

To compute HFS constants between the ground and excited states a direct multireference configuration interaction approach was used accounting for spin-orbit effects (SODCI) \cite{Alekseyev:04a} and employing the spin-orbit selection procedure \cite{Titov:01}  for configuration state functions.
13 electrons ($6s^26p^2$ for Pb and $1s^22s^22p^5$ for F) were explicitly treated in this calculation. The outer-core electrons of Pb were frozen as spinors \cite{Titov:99} using the GRECP technique.

\section{Molecular Hamiltonian}

We represent the molecular Hamiltonian as

\begin{equation}
{\rm \bf H}_{\rm mol} = {\rm \bf H}_{\rm rot} + {\rm \bf H}_{\rm hfs} + {\rm \bf H}_{1}.
\end{equation} 
Here ${\rm \bf H}_{\rm rot}$ is the rotational Hamiltonian and ${\rm \bf H}_{\rm hfs}$ is the hyperfine interaction between electrons and nuclei. ${\rm \bf H}_{1}$ includes the nuclear spin -- rotational interaction and also effectively takes into account the ${\rm \bf H}_{\rm rot}$ and ${\rm \bf H}_{\rm hfs}$ interactions between $^2\Pi_{\pm1/2} $ and other electronic states. For ${\rm \bf H}_{1}$ we have

\begin{equation}
{\rm \bf H}_{1} = c_1{\bf I}_1\cdot{\bf J} + c_2{\bf I}_2\cdot{\bf J} + c_0{\bf I}_1\cdot{\bf I}_2
\end{equation}
The parameters required to calculate the matrix elements of ${\rm \bf H}_{\rm rot}$ and ${\rm \bf H}_{\rm hfs}$ in the basis set (\ref{basis}) are determined by equations (\ref{rot}) - (\ref{Aperp2}):

\begin{eqnarray}
E_J = B\cdot J(J+1) - D\cdot [J(J+1)]^2\ ,
\label{rot}
\end{eqnarray}

\begin{eqnarray}
p_J = \frac{\hbar^2}{M_rR^2}\langle\chi_J\langle\Psi_{^2\Pi_{1/2}}|J^e_+|\Psi_{^2\Pi_{-1/2}}\rangle
\chi_J\rangle\ ,
\label{double}
\end{eqnarray}

\begin{eqnarray}
{\rm where}~~ p_J \equiv {\rm p} + {\rm p}_{\rm D}\cdot J(J+1)\ ,
\label{double1}
\end{eqnarray}

\begin{eqnarray}
\tilde{p} = 2B\langle\Psi_{^2\Pi_{3/2}}|J^e_+|\Psi_{^2\Pi_{1/2}}\rangle\ ,
\label{double2}
\end{eqnarray}

\begin{multline}
\label{Apar}
A_{J \parallel n} = \\
  \frac{\mu_{\rm F(Pb)}}{I\Omega}\langle\chi_J 
   \langle
   \Psi_{^2\Pi_{1/2}}|\sum_i\left(\frac{\bm{\alpha}_i\times
\bm{r}_i}{r_i^3}\right)
_z|\Psi_{^2\Pi_{1/2}}\rangle \chi_J\rangle\ ,
\end{multline}
\begin{eqnarray}
{\rm where}~~ A_{J \parallel n} \equiv  \rm{A}_{\parallel n} + \rm{A}_{\rm D\parallel n}\cdot J(J+1)\ ,
\label{Apara}
\end{eqnarray}

\begin{multline}
\label{Aperp1}
A_{J \perp n} = \\
  \frac{\mu_{\rm F(Pb)}}{I}\langle\chi_J 
   \langle
   \Psi_{^2\Pi_{1/2}}|\sum_i\left(\frac{\bm{\alpha}_i\times
\bm{r}_i}{r_i^3}\right)
_+|\Psi_{^2\Pi_{-1/2}}\rangle \chi_J\rangle\ ,
\end{multline}
\begin{eqnarray}
{\rm where}~~ A_{J \perp n} \equiv  \rm{A}_{\perp n} + \rm{A}_{\rm D \perp n}\cdot J(J+1)\ , {\rm and}
\label{Aperp1a}
\end{eqnarray}

\begin{eqnarray}
\tilde{A}_{\perp n} =
   \frac{\mu_{\rm F(Pb)}}{I}
   \langle
   \Psi_{^2\Pi_{3/2}}|\sum_i\left(\frac{\bm{\alpha}_i\times
\bm{r}_i}{r_i^3}\right)
_+|\Psi_{^2\Pi_{1/2}}\rangle\ .\qquad
\label{Aperp2}
\end{eqnarray}
Here ${\rm p}_{\rm D}$, $\rm{A}_{\rm D\parallel n}$,  and $\rm{A}_{\rm D \perp n}$ are centrifugal corrections to the corresponding parameters, ${\rm p}$, $\rm{A}_{\parallel n}$, and  $\rm{A}_{\perp n}$. Note that
 $\rm n{=}1$ for F and $\rm n{=}2$ for $^{207}$Pb.
To evaluate the vibration-rotation energies, $E_J$, and wavefunctions, $\chi_J$,  the Schr{\"o}dinger equation (\ref{Shr}) was solved for the ground vibration level and several lowest rotational levels numerically using the MOLCAS 4.1 package \cite{MOLCAS}.
  The potential energy values calculated for eight internuclear distances, $R= 3.3, 3.5, 3.7, 3.8, 3.9, 4.0, 4.2, 4.3 $ a.u., were used. Then parameters $B$ and $D$ were obtained by fitting $E_J$ using eq.~(\ref{rot}).
  To compute
${\rm p}$, ${\rm p}_{\rm D}$, $\rm{A}_{\parallel(\perp) n}$ and $\rm{A}_{\rm D \parallel(\perp) n}$ we first calculated corresponding properties for the distances listed above. Then the $J$-dependent matrix elements (\ref{double}), (\ref{Apar}) and (\ref{Aperp1}) over the $\chi_J$ wavefunctions were evaluated and the ${\rm p}$, ${\rm p}_{\rm D}$, $\rm{A}_{\parallel(\perp) n}$,
$\rm{A}_{\rm D \parallel(\perp) n}$ parameters were fitted using eqs.~(\ref{double1}), (\ref{Apara}) and (\ref{Aperp1a}). Parameters $\tilde{p}$ and $\tilde{A}_{\perp n}$ (eqs.~(\ref{double2}) and (\ref{Aperp2})) were computed using only the electronic wavefunction calculated at $R=3.9$~a.u..

\section{Results and discussion}

\begin{table*}
\caption{Parameters used to obtain the pure rotational spectra for $^{208}$Pb$^{19}$F}
\begin{tabular}{ccccccc}
fit & 1 & 2 & 3 & 4 & 5 & 6 \\
\hline
  B (cm$^{-1}$)           &  0.228027888 &  0.228034349      &    0.228034349       &  0.228034349 & 0.228027888           &0.228027888            \\  
D$\times 10^{7}$ (cm$^{-1}$)&  1.855      &   1.855         &     1.855         &      1.855         & 1.856               &  1.856                \\      
  p (cm$^{-1}$)            &  -0.138214051&  -0.138214052    &     -0.138214052     &   -0.138214042     & -0.138214041    & -0.138214041          \\
p$_{\rm D}\times 10^{7}$(cm$^{-1}$)     &  -1.020   &    -1.020      &      -1.020       &   -1.037        & -1.037        &  -1.037               \\
$\tilde{p}(cm^{-1})$& {\bf fixed to zero} &    0.4625  &      0.4625          &    0.4625 & {\bf fixed to zero }                      &{\bf fixed to zero}      \\               
 A$_{\parallel 1}$(MHz)   &   409.906     &    409.906        &       409.919        &   409.919 & 409.906                 &   409.905             \\        
 A$_{\rm D{\parallel 1}}\times 10^{4}$(MHz)  &{\bf fixed to zero} & {\bf fixed to zero} &  {\bf fixed to zero }     &   -1.3  & -2.9        &{\bf fixed to zero}    \\
 A$_{\perp 1}$(MHz)       &    255.9936   &   255.9935        &      255.9935        &   255.9912  & 255.9911              &    255.9911           \\     
 A$_{\rm D \perp 1}\times 10^{4}$(MHz) & {\bf fixed to zero }& {\bf fixed to zero }  &  {\bf fixed to zero }     &    5.2    &  5.3        &   5.3                 \\
 c$_1$(MHz)               &   0.0095      &    0.0095         &     0.0051           &    0.0051    & 0.0095               &   0.0095              \\ 
$\tilde{A}_{\perp 1}$(MHz) & {\bf  fixed to zero} & {\bf  fixed to zero } &    157               &   157  & {\bf fixed to zero }           &{\bf fixed to zero}        
\end{tabular}
\label{param208}
\end{table*}

\begin{table*}
\caption{Parameters used to obtain the pure rotational spectra for $^{207}$Pb$^{19}$F }
\begin{tabular}{ccccc}
fit & 1 & 2 & 3 &  calc.  \\
\hline
  B (cm$^{-1}$)                              & 0.228126407 & 0.228119947     &0.228119945               &  0.229       \\ 
D$\times 10^{7}$ (cm$^{-1}$)                 & 1.865       & 1.867           & 1.869                    &    1.83      \\ 
  p (cm$^{-1}$)                              &-0.138270134 & -0.138270134    & -0.138270110             & -0.140       \\ 
p$_{\rm D}\times 10^{7}$(cm$^{-1}$)          & -1.102      & -1.101          &  -1.098                  &              \\ 
$\tilde{p}(cm^{-1})$                         & 0.4625      & {\bf fixed to zero}  &{\bf fixed to zero }        &   0.368$^a$  \\ 
 A$_{\parallel 1}$(MHz)                      & 409.910     & 409.897         &  409.896                 &  437         \\ 
 A$_{\rm D{\parallel 1}}\times 10^{4}$(MHz)  & -0.2        & -5.4            &{\bf fixed to zero}       &    -4.9      \\ 
 A$_{\perp 1}$(MHz)                          & 255.9921    & 255.9920        & 255.9919                 &  232         \\ 
 A$_{\rm D \perp 1}\times 10^{4}$(MHz)       &  4.5        & 4.5             &  4.7                     &   7.1        \\ 
 c$_1$(MHz)                                  &  0.0051     & 0.0095          & 0.0095                   &              \\ 
$\tilde{A}_{\perp 1}$(MHz)                   & 153         & {\bf fixed to zero}  &{\bf fixed to zero}       & 166          
\footnote[1]{This parameter was calculated in 13-electron approximation using SODCI method (see text).} \\ 
 A$_{\parallel 2}$(MHz)                      & 10146.5957  & 10146.6691      & 10146.6697               &  9796        \\ 
 A$_{\rm D \parallel 2} \times 10^{3}$(MHz)  &  5.7        & 5.7             &{\bf fixed to zero}       &        5.8   \\ 
 A$_{\perp 2}$(MHz)                          & -7264.0375  & -7264.0375      & -7264.0381               &  -6911       \\ 
 A$_{\rm D \perp 2} \times 10^{3}  $(MHz)    &  -6.4       & -6.4            &  -6.5                    &  -5.9        \\ 
 c$_2$(MHz)                                  &   0.0617    & 0.0367          & 0.0375                   &              \\ 
$\tilde{A}_{\perp 2}$(MHz)                   &  -896       & {\bf  fixed to zero} & {\bf fixed to zero}      &  -1217$^a$   \\ 
 c$_0$(MHz)                                  &  0.0011     &  0.0013         & 0.0015                   &              
\end{tabular}
\label{param207}
\end{table*}

   As mentioned above, the
parameters of the molecular effective spin-rotational Hamiltonian for PbF were determined in the present work both by fitting the experimentally observed transitions and on the basis of {\it ab~initio} calculations of the $\Psi_{^2\Pi_{1/2}}$ and $\Psi_{^2\Pi_{3/2}}$ states. To clarify the relative role of the different parameters we have performed several fits with some parameters fixed to zero. The  parameters thus obtained are listed in Tables \ref{param208} and \ref{param207}.
The calculated parameters agree reasonably well with the fitted ones. The present $\rm{A}_{\parallel 2}=9796$~MHz and $\rm{A}_{\perp 2}=-6911$~MHz values are in a good agreement with those obtained in paper \cite{Baklanov:10}, which were found to be $9727$~MHz and $-6860$~MHz, respectively. In this paper we use a larger basis set and the FS-RCCSD method for electronic structure calculations instead of the SODCI method used in \cite{Baklanov:10}.  Besides, in the latter paper the $\rm{A}_{\parallel 2}$ and $\rm{A}_{\perp 2}$ values were evaluated using the only electronic wavefunction obtained at $R=4.0$ a.u., whereas in the present paper they are obtained from eqs.~(\ref{Apara}) and (\ref{Aperp1a}), respectively.

The deviations of our fits from the observed microwave transitions are given in Table \ref{spectr208} for  $^{208}$Pb$^{19}$F and in Table \ref{spectr207} for $^{207}$Pb$^{19}$F. The error was defined as

\begin{equation}
\sqrt{\sum_{i=1}^{N_t}{\left( \frac{\nu_{fit} - \nu_{obs}}{\delta_i} \right)^2 } },
\end{equation}
where $N_t$ is the number of measured transitions and $\delta_i$ is the experimental error for a given transition. In fit~1 for the $^{208}$Pb$^{19}$F the parameters $\tilde{p}$, $\tilde{A}_{\perp 1}$, $\rm{A}_{\rm D\parallel 1}$,  and $\rm{A}_{\rm D \perp 1}$ are fixed to zero. Note that the parameters $\tilde{p}$ and $\tilde{A}_{\perp 1}$ take into account the interactions of ${\rm \bf H}_{\rm rot}$ and ${\rm \bf H}_{\rm hfs}$, respectively, with the $^2\Pi_{\pm3/2}$ electronic state. In the cases of fit~2, fit~3 and fit~4 we sequentially include parameters $\tilde{p}$, $\tilde{A}_{\perp 1}$ and $\rm{A}_{\rm D\parallel 1}$ with $\rm{A}_{\rm D \perp 1}$. Data from Table \ref{spectr208} show that accuracy for the cases of fit~2 and fit~3 is actually the same as for fit~1. Note that the inclusion of the  $\tilde{p}$ and $\tilde{A}_{\perp 1}$ parameters strongly influence the $c_1$ parameter. This is in agreement with the fact that $c_1$, in particular,  takes into account the ${\rm \bf H}_{\rm rot}$ and ${\rm \bf H}_{\rm hfs}$ interactions between the $^2\Pi_{\pm1/2} $ and $^2\Pi_{\pm3/2}$ electronic states.
   On the other hand,
one can see that parameters $\rm{A}_{\rm D\parallel n}$ and $\rm{A}_{\rm D \perp n}$ are
important and that their inclusion allows one to reproduce transition energies within the experimental accuracy.
In fit~5 we keep  $\rm{A}_{\rm D\parallel 1}$ and $\rm{A}_{\rm D \perp 1}$ while removing the
$\tilde{p}$ and $\tilde{A}_{\perp 1}$ parameters. This only slightly influences on accuracy of the fit.

For $^{207}$Pb$^{19}$F we have performed the following fits. In fit~1 all the parameters were used. In fit~2 we have excluded the parameters corresponding to interactions with the $^2\Pi_{\pm3/2}$ electronic states. Data from Table \ref{spectr207} show that this does not influence the accuracy of the fit.

In fits~1 and 2 for $^{207}$Pb$^{19}$F and in fits~4 and 5 for $^{208}$Pb$^{19}$F we have included centrifugal corrections for both $\rm{A}_{\rm \parallel n}$ 
and
for $\rm{A}_{\rm \perp n}$
parameters.
However, it is clear from general theory (see eq. (2) in Ref. \cite{Shafer-Ray:08E} and eq. (2) in Ref. \cite{Sauer:99}) 
that for a $^2\Pi_{\pm1/2}$ state $\rm{A}_{\perp n}$ is the leading term, and  the $J-$dependence of the hyperfine splitting shows that 
the contribution due to $\rm{A}_{\parallel n}$ is further reduced as $J$ increases. Of course the centrifugal corrections 
become more important at higher $J$, where the relative influence of the $\rm{A}_{\parallel n}$ terms is quite diminished. Hence the 
inclusion of the $\rm{A}_{\rm D\parallel n}$ centrifugal correction parameters should be less important to the fits than 
 $\rm{A}_{\rm D \perp n}$. 
%
To confirm this we have
performed fit~6 for $^{208}$Pb$^{19}$F and fit~3 for $^{207}$Pb$^{19}$F where we have excluded the $\rm{A}_{\rm D\parallel n}$ parameters.
Data in Tables \ref{spectr208} and \ref{spectr207} show that this does not influence the accuracy of the fits.

In paper \cite{Mawhorter:11}, new $d_{c1}$ and $d_{c2}$ parameters were introduced to fit the spectra of PbF. Two possibilities were given for their origin. Our parameters $\rm{A}_{\rm D \perp 1}$ and $\rm{A}_{\rm D \perp 2}$ are also required to fit the spectra of PbF and 
are in robust agreement within experimental uncertainty
with $d_{c1}$ and $d_{c2}$, which were found to be $.00056(10)$~MHz and $-.007(2)$~MHz, respectively. Therefore, we suggest that parameters $d_{c1,2}$ can be
   rather treated
as centrifugal corrections to the $\rm{A}_{\perp 1,2}$ parameters in accordance with the second interpretation given in \cite{Mawhorter:11}.

\begin{table}
\caption{  Observed pure rotational transition frequencies $\nu_{obs}$ \cite{Mawhorter:11}
of the X$_1$ state of $^{208}$Pb$^{19}$F. The deviation of n-th fit  is given by $\Delta_n= \nu_{fit} - \nu_{obs}$}
\begin{tabular}{cccccccccc}
  levels   &    $\nu_{obs}$ & $\Delta$ \cite{Mawhorter:11} & $\Delta_1$  &$\Delta_2$   &    $\Delta_3$   &     $\Delta_4$    &   $\Delta_5$  &   $\Delta_6$ \\
\hline
 1-3       &  3922.5065(20) &     -6       &  -10    & -10       &     -9      &      -5  &  -5 & -3\\
 2-4       &  4194.7773(7)  &     -1       &   0  &  0       &     0       &      3       &   3 & 1 \\
 1-4       &  4229.7176(30) &      15      &  29  & 29       &     29      &      19      &  20 & 19 \\
 6-7       &  8117.3017(10) &      -8      &  -6  & -5       &     -5      &      -5      &  -5 & -5 \\
 5-7       & 8199.8478(9)   &       4      &  6   & 6       &     7       &       6       &  6  & 5 \\
 6-8       & 8307.5180(20)  &      -2      &  8   & 9       &     8       &       6       &  6  & 5 \\
 5-8       & 8390.0664(20)  &      -12     &  -3  & -3       &    -3       &      -7      &  -7 & -7 \\
10-11      & 12277.6822(7)  &       1      &  12  & 13       &    13       &       3      &   3 & 3 \\
 9-12      & 12540.8465(8)  &      -5      & -17  & -17       &    -17      &       -3    &  -3 & -3 \\
14-15      & 16428.5160(10) &       2      &  11  & 11       &    11       &      -2      &  -2 & -2 \\
13-16     & 16688.4929(20) &       17     &   -35 & -36      &    -36      &        10    &  10 & 11 \\
 1-5       & 18414.5880(5)  &       1      &   5  & 5        &    5        &       2      &  2  & 2\\
 2-6       & 18462.1933(5)  &       2      &  -7  & -7       &   -8        &       0      &  0  & -1 \\
 1-6       & 18497.1352(5)  &       2      &   6  &  5        &    6        &       1     &  1  & 1 \\
 4-7       & 22384.7171(5)  &       1      &  -7  & -7        &   -6        &      -2     &  -2 &  -1 \\
 4-8       & 22574.9344(5)  &       -2     &  -3  & -3        &   -3        &      -1     &  -1 &  0\\
 3-7       & 22691.9306(5)  &       -2     &   8  & 8         &   7         &       -1    &  -1 &  -3 \\
error      &                &      1.86    &  4.84    & 4.83      &   4.82      &      1.60 &  1.62 & 1.63  \\
\hline
\end{tabular}
\label{spectr208}
\end{table}

\begin{table}
\caption{  Observed pure rotational transition frequencies $\nu_{obs}$ \cite{Mawhorter:11}
of the X$_1$ state of $^{207}$Pb$^{19}$F. The deviation of $n$-th
fit  is given by $\Delta_n= \nu_{fit} - \nu_{obs}$}
\begin{tabular}{ccccccccc}
  levels   &    $\nu_{obs}$ & $\Delta$ \cite{Mawhorter:11} & $\Delta_1$  &$\Delta_2$  &$\Delta_3$  \\
\hline
3-6   & 3187.4875(20) & 5      & -6  &  -6   & -7 \\
2-6   & 3219.8137(7) & -3      &  1 &  1     & -1 \\
10-11 & 4455.4540(25) & 21     & 16  &  16   & 14 \\
9-12  & 4699.2265(25) & 29     & 29  & 29    & 26 \\
1-4   & 8495.0022(7) & 1       &  5 &  5     &  1 \\
18-19 & 8620.5475(10) & -8     & 20  & 20    & 18 \\
1-5   & 8687.2098(7) & 6       & 4  & 4      & 1\\
8-13  & 11 682.5211(7) & 6     & 2  & 2      & -2\\
7-13  & 11 715.3703(7) & -14   & -13  & -13  & -17\\
8-14  & 11 867.6415(10) & -8   & -9  & -8    & -12 \\
7-14  & 11 900.4870(7) & 9     &  13 & 14    & 10 \\
3-7   & 14 430.1830(5) & 13    &  2 & 2      & 2 \\
2-7   & 14 462.5104(5) & -8    & -3  & -3    & -4\\
3-8   & 14 463.0311(5) & 3     & -2  & -2    & -2\\
2-8   & 14 495.3580(5) & -11   & -2  &  -2   & -3\\
16-21 & 15 865.1888(5) & -8    & -6  & -5    & -4\\
15-22 & 16 108.7271(5) & -4    & -2  & -2    & 0\\
2-9   & 18 333.5013(5) & 3     &  5 &  6     & 7\\
3-10  & 18 380.8711(5) & 5     &  1 &  2     & 0\\
2-10  & 18 413.1982(5) & -12   &  -1 & -1    & -3\\
5-11  & 22 377.8342(5) & 2     & 3  &  4     & 5\\
5-12  & 22 541.9123(5) & -1    & -2  & -1    & 2\\
4-11  & 22 570.0427(5) & -2    & -6  & -5    & -5\\
1-7   & 22 658.9018(5) & -5    & -7  & -6    & -6\\
1-8   & 22 691.7486(5) & -2    & 3  &  3     & 3\\
6-13  & 22 958.0652(5) & -1    & 1  & 1      & -2\\
6-14  & 23 143.1846(5) & -4    & 1  & 1      & -2\\
5-13  & 25 687.0601(5) & -4    & -2  & -1    & -2\\
5-14  & 25 872.1789(5) & -1    &  4 & 4      & 4\\
4-13  & 25 879.2667(5) & 12    &  8 & 8      & 8\\
4-14  & 26 064.3873(5) & -3    & -3  & -4    & -4\\
error &                &  6.68     & 5.12  & 5.13 & 5.13\\
\hline
\end{tabular}
\label{spectr207}
\end{table}


This work is supported by the SPbU Fundamental Science Research grant from Federal budget No.~0.38.652.2013 and RFBR Grant No.~13-02-01406. L.S.\ is also grateful to the Dmitry Zimin ``Dynasty'' Foundation. The molecular calculations were performed at the Supercomputer ``Lomonosov''. R.J.M.\ would like to acknowledge support from Pomona College and the German Academic Exchange Service (DAAD).


\end{document}